\input epsf
\documentclass[nofootbib,aps,twocolumn,10pt,preprint,floatfix]{revtex4}

\usepackage{graphicx}

\begin{document}
\preprint{WSU-HEP-0704}

\title{On the Non-Manifest Left-Right Model Contribution
to the
Lifetime Difference in $D^0 - \bar{D}^0$ Mixing.}

\author{G. K. Yeghiyan}

\email{ye_gagik@wayne.edu}

\affiliation{\vskip 0.3cm Department of Physics and Astronomy \\
Wayne State University, Detroit MI, USA.\vskip 0.05cm}

\abstract{\vskip 0.3cm New physics contribution to the lifetime
difference in $D^0 - \bar{D}^0$ mixing is re-examined within the
non-manifest Left-Right Symmetric Model. Diagrams with one of
$\Delta C = 1$ transitions, mediated by a propagator with $W_L -
W_R$ mixing, are revisited. While these diagrams are believed
to give the dominant contribution, compatible with the experimental
data, it is shown that due to GIM cancelation, such diagrams are
negligible in sum. Thus, Left-Right Symmetric Model contribution to
the lifetime difference in $D^0 - \bar{D}^0$ mixing is about two
orders of magnitude less than actual experimental value for
$\Delta \Gamma_D$.}}

\maketitle

\vspace{0.5cm}

Observation of $D^0 - \bar{D}^0$ oscillations by BaBaR and Belle
collaborations \cite{10,11} has revived the theoretical interest
to this phenomenon. In particular, new physics contribution to
$D^0 - \bar{D}^0$ mixing has been re-considered in several
publications \cite{12} - \cite{1}.

The impact of a new physics on the lifetime difference in
$D^0 - \bar{D}^0$ mixing has been in details studied in \cite{3}.
Model independent analysis has been performed and the derived
analytical formulae have been then applied within several
extensions of the Standard Model. It has been shown that
to the lowest order in the perturbation theory, new
physics contribution to $\Delta \Gamma_D$ may be several
orders of magnitude greater than that of the Standard Model.
Later on the lifetime difference in $D^0 - \bar{D}^0$ mixing
has been also considered in \cite{19,25} and \cite{1},
within the supersymmetric
models with R-parity violation and the Left-Right Symmetric
Model, respectively.

\begin{figure}[h]
\epsfxsize=6cm
\epsfysize=4cm
\hspace{0.5cm}
\epsffile{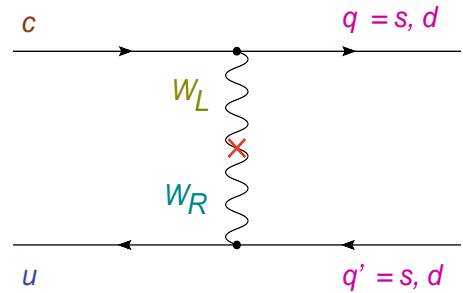}
\caption{$\Delta C = 1$ transition mediated by a propagator
with $W_L - W_R$ mixing.}
\label{f1}
\end{figure}

It has been argued in
\cite{1} that within the non-manifest Left-Right (LR) Symmetric
Model, new physics contribution to the lifetime difference in $D^0 -
\bar{D}^0$ mixing may be significant:
\begin{equation}
|y_{LR}| \equiv \frac{|\Delta \Gamma_{D_{LR}}|}{2 \Gamma_D}
\leq 1.4 \times 10^{-3}, \label{1}
\end{equation}
which means that $y_{LR}$ may be of the same order as the
experimental value of $y$ \cite{2},
\begin{equation}
y \equiv \frac{\Delta \Gamma_D}{2 \Gamma_ D} = (
6.6 \pm 2.1) \times 10^{-3} \label{2}
\end{equation}
This result has been derived by considering the box diagrams
with one of $\Delta C = 1$
transitions being generated by a new physics (NP) interaction and
mediated by a propagator with $W_L - W_R$ mixing
(Fig.~\ref{f1}).
Note that $W_R$ part of the
propagator couples with the u-quark, which allows one to
remove a power of the suppression in terms of
$\lambda=\sin{\theta_C} \approx 0.23$.

In this letter we revisit the contribution of the box diagrams with
the new physics generated $\Delta C = 1$ transition, presented in
Fig~\ref{f1}. While the analysis of ref. \cite{1} is restricted by
considering only the diagrams with the intermediate s-quark states,
i.e. $q = s$ and $q^\prime = s$, we include also the diagrams with
$q = d$ and/or $q^\prime = d$. We will see that diagrams with the
intermediate d-quark states may not be neglected, in spite of $m_d
\ll m_s$. Moreover, they play crucial role in taking into account
GIM cancelation effects properly.

In this paper we show that box diagrams
with the new physics generated
$\Delta C = 1$ transition, presented
in Fig~\ref{f1}, are negligible {\it in sum} due to
GIM cancelation. Thus,
one must replace the bound on $y_{LR}$, given by
equation (\ref{1}), by
\begin{equation}
|y_{LR}| \leq 8.8 \times 10^{-5} \label{3}
\end{equation}
This constraint on $y_{LR}$ has been derived in
\cite{3}, neglecting $\Delta C =1$ transition
presented in Fig.~\ref{f1}.

For $\Delta C = 1$ interaction in Fig.~\ref{f1},
the relevant part of the low-energy effective
Hamiltonian has the
following form:
\begin{eqnarray}
\nonumber
H_{W_{L-R}}^{\Delta C = 1} = - \frac{4 G_F \xi_g }{\sqrt{2}}
\sum_{q, q^\prime} V_{cq}^{* L} V_{uq^\prime}^R
\Big[\bar{C}_1(m_c) Q_1 + \\ + \bar{C}_2(m_c) Q_2 \Big]
\label{14} \\
\nonumber
Q_1 = \bar{u}_i \gamma^\nu P_R q^\prime_j
\bar{q}_j \gamma_\nu P_L c_i, \hspace{0.3cm}
Q_2 = \bar{u}_i \gamma^\nu P_R q^\prime_i
\bar{q}_j \gamma_\nu P_L c_j
\end{eqnarray}
where $P_L = (1 - \gamma_5)/2$, $P_R = (1 + \gamma_5)/2$,
$i$, $j$ stand for color indices, $V^L$, $V^R$ are the left- and
right-handed quark CKM matrices and $\xi_g$ is defined
in \cite{1}.
If only one $\Delta C=\nolinebreak1$ transition in the box diagrams
is generated by an NP interaction, the approach
described in ref. \cite{3} may be used.
For the new physics $\Delta C = 1$ effective Hamiltonian given by equation (\ref{14}),
it is not hard to see that only the term
$I_4(x_q, x_{q^\prime})~\langle~\bar{D}^0~|~O_4^{ijkl}~|~D^0~\rangle$ in
equation~(7) of \cite{3} contributes. Basically, this result is in
agreement with that of ref.~\cite{1}, however there is an essential difference.
While $q = q^\prime = s$ in \cite{1}, we take here $q = s, d$ and
$q^\prime = s, d$.
If one denotes by $y_{LR}^{(1)}$ the considered here
contribution to the
lifetime difference in $D^0 -
\bar{D}^0$ mixing, then, using eqs. (7), (9), (10) in ref. \cite{3} (setting  
there
$D_{q q^\prime} = - \left(G_F/\sqrt{2}\right) \xi_g V_{cq}^{L*} V_{uq^\prime}^R$,
$\bar{\Gamma}_1 = \gamma^{\nu} P_R$,
$\bar{\Gamma}_2 = \gamma_{\nu} P_L$),
it is straightforward to show after doing some algebra that
\begin{equation}
y_{LR}^{(1)} \ = \sum_{q, \ q^\prime} C_{LR}^{q q^\prime}
\ V_{c q^\prime}^{L^*}
V_{u q^\prime}^R \ \left[ K_2 \langle Q^\prime \rangle +
K_1 \langle \tilde{Q}^\prime \rangle \right] \label{4}
\end{equation}
where
\begin{eqnarray}
\nonumber
C_{LR}^{q q^\prime} \ = \ \frac{G_F^2 m_c^2 \ \xi_g}{2 \pi
m_D \Gamma_D} \ V_{c q}^{L^*}
V_{u q}^L \ \sqrt{x_{q^\prime}} \ \Big[(1 - x_{q^\prime})^2
- \\
- 2 x_q x_{q^\prime} - x_q^2 \Big] \label{5}
\end{eqnarray}
and the notations in (\ref{4}) and (\ref{5}) are the same as in
\cite{1}.

Formulae (\ref{4}) and (\ref{5}) are generalization of formulae (3)
and (4) of ref. \cite{1} for the case when both s- and d-quark
intermediate states are considered, thus $C_{LR}$ of \cite{1} is
replaced here by $C_{LR}^{q q^\prime}$ and sum over q, $q^\prime$ is
implemented. Else, in order to take properly into account GIM cancelation
effects, we keep in equation (\ref{5}) higher order
terms in the expansion in powers of $x_q \equiv m_q^2/m_c^2$ and
$x_{q^\prime} \equiv m_{q^\prime}^2/m_c^2$.

It is worth to note that dependence on $x_q$ appears only in the
next-to-next-to-leading order terms of this expansion. The difference
in the behavior of $y_{LR}^{(1)}$ with $x_q$ and with $x_{q^\prime}$
is related to different chiralities  of the light quarks $q$
and $q^\prime$ in (\ref{14}). More detailed discussion of the behavior of
$D^0 - \bar{D^0}$ mixing amplitude with the light quark
masses, depending on these quarks chiralities, may be found in refs.
\cite{22}-\cite{24}. Discussion for a particular case of the width
difference is also available in \cite{21, 17}.

It is clear from (\ref{5}) that
if one takes the limit $x_d \equiv m_d^2/m_c^2 = 0$,
$C_{LR}^{q q^\prime} = 0$ for $q^\prime = d$. Thus, formula
(\ref{4}) is significantly simplified:
\begin{equation}
y_{LR}^{(1)} \ = \left[C_{LR}^{s s} + C_{LR}^{d s} \right]
\ V_{c s}^{L^*}
V_{u s}^R \ \left[ K_2 \langle Q^\prime \rangle +
K_1 \langle \tilde{Q}^\prime \rangle \right] \label{6}
\end{equation}
where
\begin{eqnarray}
&& \hspace{-1.5cm}
C_{LR}^{s s} \ = \ \frac{G_F^2 m_c^2 \ \xi_g}{2 \pi
m_D \Gamma_D} \ V_{c s}^{L^*}
V_{u s}^L \ \sqrt{x_s} \ \Big[(1 - x_{s})^2
-3 x_s^2 \Big] \label{7} \\
&& \hspace{-1.5cm}
C_{LR}^{d s} \ = \ \frac{G_F^2 m_c^2 \ \xi_g}{2 \pi
m_D \Gamma_D} \ V_{c d}^{L^*}
V_{u d}^L \ \sqrt{x_s} \ (1 - x_{s})^2 \label{8}
\end{eqnarray}

As it follows from (\ref{6}) - ({\ref{8}),
in the limit $m_d = 0$ there is an
additional contribution - as compared to that of ref.
\cite{1} -  from the diagram in Fig.~\ref{f1}
when $q = d$ and $q^\prime = s$:  $C_{LR}^{d s}
\neq 0$. Moreover,
using the fact that $V_{c s}^{L^*}
V_{u s}^L \approx - V_{c d}^{L^*}
V_{u d}^L + O(\lambda^5)$, it is not hard to see that
$C_{LR}^{s s} \approx - C_{LR}^{d s}$ with
accuracy of the terms $\sim \lambda^5$ or
$\sim x_s^{5/2}$.
Thus, sum of $C_{LR}^{s s}$ and $C_{LR}^{d s}$ is much less in
the absolute value than these quantities by themselves.
This is manifestation of (approximate) GIM cancelation that
makes $y_{LR}^{(1)}$ negligible.

Using the unitarity condition,
\begin{equation}
V_{c s}^{L^*} V_{u s}^L + V_{c d}^{L^*}
V_{u d}^L + V_{c b}^{L^*} V_{u b}^L = 0 \label{9}
\end{equation}
one gets after doing some algebra
\begin{eqnarray}
\nonumber
\hspace{-0.6cm}
C_{LR}^{s s} + C_{LR}^{d s} \ = \ \frac{G_F^2 m_c^2 \ \xi_g}{2
\pi m_D \Gamma_D}  \ \sqrt{x_s} \ \Big[ - Re\left(V_{c b}^{L^*}
V_{u b}^L \right) (1 - \\
- x_{s})^2
- \ 3 \ V_{c s}^{L^*}
V_{u s}^L \ x_s^2 \Big] \label{10}
\end{eqnarray}
Note that unlike CKM products in (\ref{4}) - (\ref{8}),
$V_{c b}^{L^*} V_{u b}^L$ has non-negligible phase
\cite{4}, thus one must explicitly indicate that the real
part of this product is only relevant. It is assumed no
new source of CP-violation \cite{3} ($V_R$ is real and
no spontaneous CP-violation). In this case, the impact of
CP-violating effects on $\Delta \Gamma_D$ is negligible.

Usually, when studying $D^0 - \bar{D}^0$ oscillations, one puts
$V_{c b}^{L^*} V_{u b}^L \approx 0$, as $|V_{c b}^{L^*} V_{u b}^L|
\ll V_{c s}^{L^*} V_{u s}^L$. This is the two quark generation
mixing approximation, that is widely applied in studying $D^0 -
\bar{D}^0$ mixing effects within the Standard Model (see e.g
\cite{5}) and some of its extensions. However, in our case this
approximation is not valid. Indeed, using $Re\left(V_{c b}^{L^*}
V_{u b}^L \right) \approx A^2 \lambda^5 \rho$  and $V_{c s}^{L^*}
V_{u s}^L \approx \lambda$, it is not hard to see that the first
term in the square brackets in (\ref{10}) dominates over the last
one, for $A \approx 0.82$, $\lambda \approx 0.23$, $\rho \approx
0.23$ \cite{4} and $x_s \equiv m_s^2(m_c)/m_c^2(m_c) \approx 0.007$
\cite{25}.

To the lowest order in the perturbation theory,
one gets a rough estimate of the effect rather than a precise
numerical evaluation. In what follows, one may to
a good approximation disregard the
subdominant terms in (\ref{10}).
Then, one may rewrite equation (\ref{6}) in a more compact form:
\begin{equation}
y_{LR}^{(1)} \ = - \bar{C}_{LR} \ V_{c s}^{L^*}
V_{u s}^R \ \left[ K_2 \langle Q^\prime \rangle +
K_1 \langle \tilde{Q}^\prime \rangle \right] \label{11}
\end{equation}
where
\begin{equation}
\bar{C}_{LR} \ = \ \frac{G_F^2 m_c^2 \ \xi_g}{2 \pi
m_D \Gamma_D}  \ Re\left(V_{c b}^{L^*} V_{u b}^L \right) \
\sqrt{x_s}  \label{12}
\end{equation}

We parameterize $\langle Q^\prime \rangle$ and
$\langle \tilde{Q}^\prime \rangle$, using the moderate
vacuum saturation approach \cite{13}:
\begin{eqnarray}
\nonumber
&&\langle Q^\prime \rangle \equiv \langle \bar{D}^0 | \bar{u}_i
\gamma^\mu P_L c_i \bar{u}_j \gamma_\mu P_R c_j |D^0 \rangle = \\
&& = - \frac{1}{2} f_D^2 m_D^2 B_D  -
\frac{1}{3} f_D^2 m_D^2 \left(\frac{m_D}{m_c}\right)^2 B_D^S
\label{15} \\
\nonumber
&&\langle \tilde{Q}^\prime \rangle \equiv \langle \bar{D}^0 | \bar{u}_i
\gamma^\mu P_L c_j \bar{u}_j \gamma_\mu P_R c_i |D^0 \rangle = \\
&& = - \frac{1}{6} f_D^2 m_D^2 B_D  -
f_D^2 m_D^2 \left(\frac{m_D}{m_c}\right)^2 B_D^S
\label{16}
\end{eqnarray}
where $f_D \approx 0.22$GeV \cite{7},
$B_D \approx 0.8$ \cite{6}, and we choose $B_D^S \approx B_D$. Then,
using $G_F = 1.166 \times 10^{-5}GeV^{-2}$,
$\Gamma_D \approx 1.6 \times 10^{-12}$GeV,
$m_D \approx 1.865$GeV,
$m_c \equiv m_c(m_c) \approx 1.25$GeV \cite{4},
$K_1 \equiv 3 C_1 \tilde{C}_1 + C_1 \tilde{C}_2 + C_2 \tilde{C}_1
\approx 3 C_1^2 + 2C_1 C_2$,
$K_2 \equiv C_2 \tilde{C}_2 \approx C_2^2$, $C_1(m_c) = -0.411$,
$C_2(m_c) = 1.208$ \cite{5}, $V_{cs}^L \approx 1 - \lambda^2/2$,
$V^R_{us} \approx 1$ and \cite{1,8} $\xi_g \leq 0.033$,
one gets
\begin{equation}
y_{LR}^{(1)} \leq 2.6 \times 10^{-7} \label{13}
\end{equation}

Thus, due to GIM cancelation,
box diagrams with the new physics generated $\Delta C = 1$ transition,
presented in Fig.~\ref{f1}, give {\it in sum} negligible contribution
to the lifetime difference in $D^0 - \bar{D}^0$ mixing.

It is left for a reader to verify that one gets negligible contribution
to $y_{LR}$ also in the case when $W_L - W_R$ propagator in Fig.~\ref{f1}
is flipped so that $W_R$ couples with the charm quark.

In what follows, one should use the
result of ref. \cite{3} that has been derived neglecting
$\Delta C = 1$ transition in Fig.~\ref{f1}. In other words,
one should use
the bound on $y_{LR}$, given by equation (\ref{3}). Thus,
within the non-manifest Left-Right Symmetric Model, new physics
contribution to to the lifetime difference in $D^0 - \bar{D}^0$
mixing is rather small.

In is worth to note here that this result has been derived considering
the diagrams with only one $\Delta C =1$ transition generated by
a new physics interaction. There are also box diagrams with both
$\Delta C = 1$ transitions occurring due to NP interactions. These
diagrams have not been considered so far, as within the Left-Right
Symmetric Model they are estimated to have a small contribution
to $\Delta \Gamma_D$. On the other hand, it is still possible
that within the non-manifest
version of the LR model, there are some corners of the parameter
space with $M_{W_R}$ below 1 TeV \cite{9},
where such diagrams are perhaps non-negligible. Study of this
possibility requires detailed and careful scanning of the
parameter space of the theory, taking into account all possible
constraints, coming from $K_L - K_S$ and $B^0 - \bar{B}^0$
mass differences, as well as other phenomenological constraints.
Such a detailed analysis is out of the scope of this brief report.

In conclusion, lifetime difference in $D^0 - \bar{D}^0$
oscillations has been re-considered within the non-manifest
Left-Right Symmetric Model. It has been shown that, due to
GIM cancelation effects, new physics contribution to the
lifetime difference in $D^0 - \bar{D}^0$ mixing is
rather small, as compared to the experimental value of
$\Delta \Gamma_D$. \\

Author is grateful to A.~A.~Petrov and Chuan-Hung~Chen for valuable
discussions. This work has been supported by the grants
NSF~PHY-0547794 and DOE~DE-FGO2-96ER41005.

\end{document}